\documentclass[aps,prx,amsmath,amssymb,twocolumn,longbibliography,floatfix]{revtex4-2}

\usepackage[utf8]{inputenc}
\usepackage[T1]{fontenc}
\usepackage{graphicx}
\usepackage{bm}
\usepackage{mathtools}
\usepackage{hyperref}
\usepackage{physics}
\usepackage{orcidlink}
\usepackage{xcolor}
\usepackage{algorithm}
\usepackage{algorithmic}

\hypersetup{colorlinks=true,linkcolor=blue,citecolor=blue,urlcolor=blue}

\newcommand{\STS}{\mathrm{STS}}

\newcommand{\Spost}{S_{\mathrm{post}}}
\newcommand{\shell}{j}

\newcommand{\preq}{p_{\mathrm{req}}}
\newcommand{\Tstar}{T^{\star}}
\newcommand{\betaann}{\beta_{\mathrm{ann}}}
\newcommand{\betarand}{\beta_{\mathrm{rand}}}
\newcommand{\pexc}{p_{\mathrm{exc}}}

\begin{document}

\title{Shots-to-Approximate-Solution Scaling in Neutral-Atom
Quantum Optimization}

\author{Junwoo Jung\orcidlink{0009-0000-4310-1598}}
\author{Jaewook Ahn\orcidlink{0000-0002-5837-6372}}
\email{jwahn@kaist.ac.kr}
\address{Department of Physics, KAIST, Daejeon 34141, Republic of Korea}
\date{\today}

\begin{abstract}\noindent
Whether neutral-atom quantum optimization protocols exhibit genuine
concentration toward low-energy solution structure remains an open question.
Here, we introduce a shots-to-approximate-solution metric, $\STS(r)$, where
$r$ denotes the approximation ratio, and evaluate it using postprocessed
outputs modeled by a degeneracy-weighted shell distribution governed by a
single effective parameter, $\beta$, that quantifies concentration toward
near-optimal independent sets.
To extract the genuine concentration effect in the quantum data, we apply
identical postprocessing to both experimental bitstrings and randomly
generated bitstrings with matched excitation density, thereby constructing an
excitation-matched random baseline.
Experiments on programmable Rydberg-atom arrays with system sizes up to 125
sites show that quantum annealing consistently exceeds the random baseline,
demonstrating enhanced concentration toward low-energy solution structure
beyond what can be attributed solely to excitation density.
The results further reveal two distinct target-dependent regimes. For
near-exact targets with $r\approx 1$, the required shot count grows
exponentially with system size and is reduced at the same exponential level by
quantum annealing within the shell-model description. By contrast, for relaxed
targets, the shot cost becomes effectively constant, and the corresponding
quantum enhancement diminishes, with the classical postprocessing
heuristic alone reaching the target in order-unity attempts. Together, these
results establish an operational method for quantifying quantum optimization
performance and clarify the regimes under which quantum approaches can yield
practical benefits.
\end{abstract}

\maketitle

\begin{figure*}[t]
  \centering
  \includegraphics[width=\linewidth]{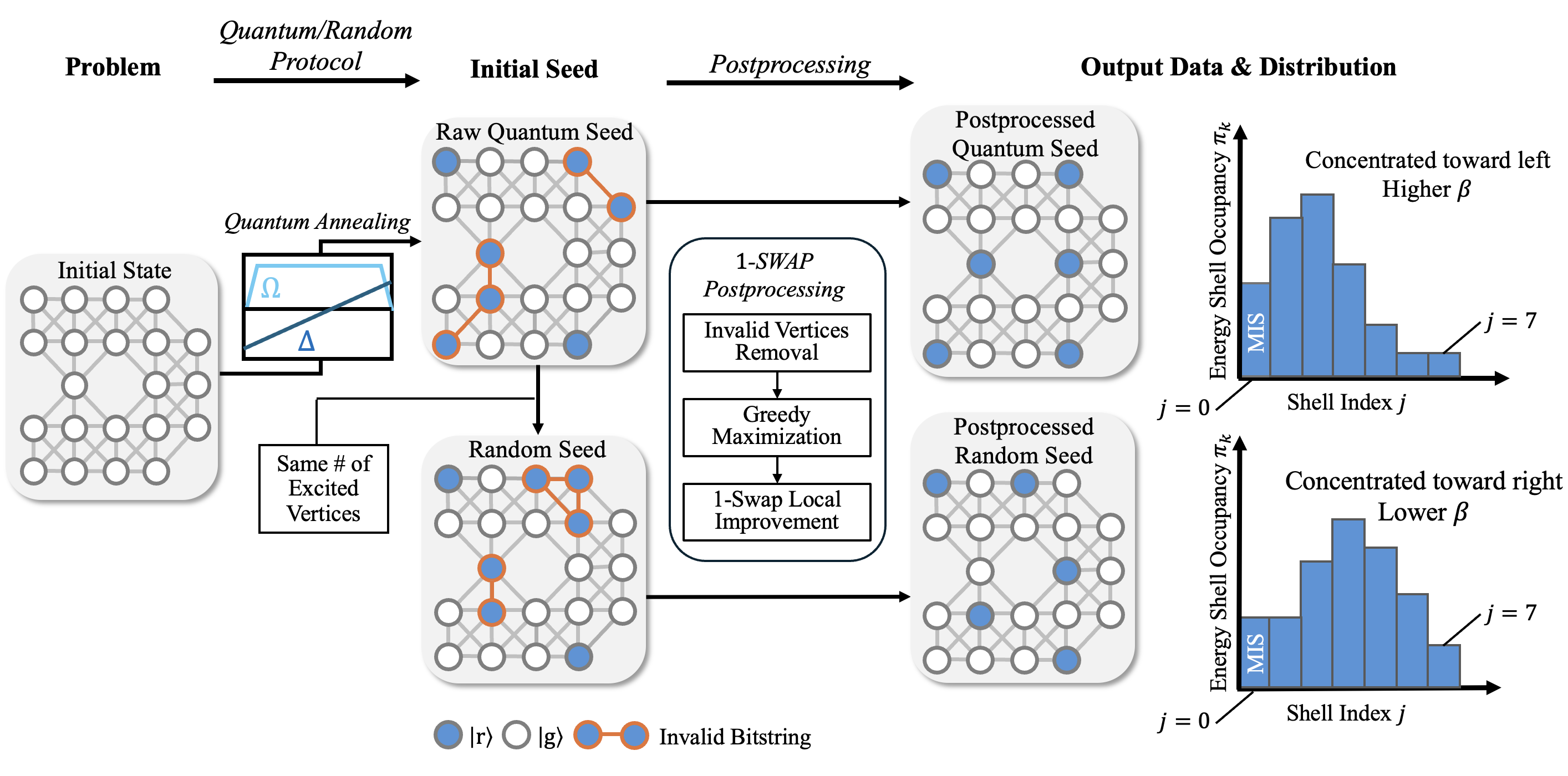}
  \caption{\textbf{Density-controlled shots-to-approximate-solution workflow.}
  An initial King's-lattice atom array is evolved by a Rydberg quantum-annealing
  schedule, producing raw quantum seeds (occupation bitstrings). To construct a
  density-controlled null model, random seeds are sampled with the same mean
  excitation density as the quantum data. Both seeds are passed through the
  identical postprocessing pipeline: invalid adjacent excitations are removed,
  the independent set is greedily maximized, and a local $1$-swap improvement is
  applied. The resulting postprocessed outputs define shell distributions over
  shell index~$j$, where $j=0$ corresponds to the MIS manifold, summarized by an
  effective quality parameter~$\beta$. Larger $\beta$ indicates stronger
  concentration toward low-$j$, high-quality independent sets, while smaller
  $\beta$ indicates a broader distribution over sub-optimal shells. Comparing
  the postprocessed quantum output with the postprocessed excitation-matched
  random baseline isolates concentration beyond the trivial effect of excitation
  density. The corresponding shots-to-approximate-solution cost $\STS(r)$ is then
  computed from the fitted shell distribution for a target approximation
  ratio~$r$.}
  \label{fig:schematic}
\end{figure*}

\section{Introduction}\label{sec:intro} \noindent
Programmable quantum many-body systems offer a natural framework for
combinatorial optimization by encoding classical cost functions onto interacting
Hamiltonians and sampling low-energy configurations through controlled quantum
dynamics~\cite{Albash2018RMP,Lucas2014,Farhi2001,Kadowaki1998,Hauke2020NatPhys}.
Among the standard benchmark problems in this setting, the maximum independent
set (MIS) problem plays an important role. For a graph $G=(V,E)$, where $V$ is
the set of vertices and $E$ is the set of edges, the objective is to find the
largest subset of vertices on $V$ with no pair connected by an edge in $E$.
Determining the maximum independent set is NP-hard in
general~\cite{Garey1979,Zuckerman2007}, and even finding constant-factor
approximation is NP-hard for general graphs~\cite{Hastad1996}. In contrast, for
structured graph families such as geometric unit-disk and King's-lattice graphs,
polynomial-time approximation schemes (PTAS) exist~\cite{Hunt1998,Nieberg2005,Erlebach2005},
making the distinction between exact and approximate optimization particularly
relevant for graph families naturally implemented in Rydberg atom arrays.
The same structure makes these instances classically tractable, and we
therefore report the measured classical reference cost alongside the quantum
shot counts (Sec.~\ref{subsec:classical}).

Neutral-atom platforms implement the MIS problem natively through the Rydberg
blockade mechanism, which energetically suppresses adjacent excitations and
thereby enforces the independence constraint at the hardware
level~\cite{Saffman2010RMP,BrowaeysLahaye2020NatPhys,Ebadi2022Science,Nguyen2023PRXQ,Song2021}.
Experiments have demonstrated both small-graph MIS implementations and large
programmable-array optimization using Rydberg
atoms~\cite{Byun2022PRXQ,Kim2022NatPhys,Ebadi2022Science}, and benchmarking
studies have compared device performance with classical heuristics while
analyzing the role of graph geometry, control schedule, and
postprocessing~\cite{Andrist2023PRR,Wurtz2022Aquila,Dupont2023PRR}. However,
these studies have largely focused on exact solutions ($r=1$) or on summary
measures of approximation quality such as the mean or distribution of
per-sample approximation ratios, without explicitly resolving how the
shots-to-approximate-solution cost varies with the target approximation quality.

In this paper, we quantify how many physical shots the annealing protocol
requires to reach a target approximation ratio~$r$. To this end, we
describe the postprocessed outputs by a degeneracy-weighted shell distribution
whose single parameter---an effective inverse temperature~$\beta$---serves as
the figure of merit for concentration toward near-optimal solutions, and we
convert the measured $\beta$ into the shot cost $\STS(r)$ at any target
ratio. Instances are organized by the hardness proxy $H(G)$ introduced in
Ref.~\cite{Ebadi2022Science}, which captures solution degeneracy and local
structure within the near-optimal manifold, and hence the Hamming-shell
statistics on which the present analysis operates.

The role of the approximation ratio on shot-based performance has been explored
in gate-based quantum optimization. In particular, Ref.~\cite{Larkin2022QST}
characterized quantum approximate optimization algorithm (QAOA) performance
through the distribution of per-sample approximation ratios, and subsequent work
has shown that targeting relaxed approximation ratios can substantially reduce
the number of required circuit runs~\cite{Chernyavskiy2025}. In the context of
analog Rydberg dynamics with deterministic postprocessing, similar
approximation-dependent structure is also of interest; however, the shot-based
cost depends on the full postprocessed output distribution rather than on
circuit-level metrics, calling for a distinct framework tailored to analog Rydberg systems
with deterministic postprocessing.

A key challenge in postprocessed analog Rydberg optimization is that apparent
improvements in the low-energy quality of postprocessed outputs can stem either
from genuine concentration toward low-shell solutions or from variations in the
per-shot excitation density. Let $\pexc$ denote the mean fraction of
Rydberg-excited atoms per shot. Because the postprocessing map operates on the
raw excitation pattern, different excitation densities can yield distinct final
shell distributions even in the absence of any nontrivial structural advantage.
Consequently, a meaningful comparison of postprocessed annealing performance
requires a baseline that explicitly controls for excitation density. We therefore
introduce an \emph{excitation-matched random baseline},
$\betarand(\pexc)$, defined by applying the same postprocessing to
$\mathrm{Bernoulli}(\pexc)$ bitstrings that reproduce the mean excitation
density of the annealing data. The corresponding residual,
$\Delta\beta_{\mathrm{ann}}\equiv
\betaann-\betarand(\pexc^{(\mathrm{ann})})$, quantifies the additional
concentration toward low-shell outputs beyond what can be attributed solely to
excitation density. Because this baseline is itself a randomized greedy heuristic, it
simultaneously provides a classical reference point for the measured shot
cost (Sec.~\ref{subsec:classical}).

The second ingredient of our framework is a one-parameter description of the
postprocessed outputs. After deterministic postprocessing, each shot yields a
valid independent set $\Spost \subseteq V$ whose size can be represented by its
Hamming-shell index $\shell=\alpha-|\Spost|$, where $\alpha$ is the MIS size,
$|\Spost|$ is the size of the postprocessed independent set, and larger $\shell$
indicates a sub-optimal solution. Empirically, the resulting shell distribution
is well described by a degeneracy-weighted form characterized by an
inverse-temperature parameter~$\beta$. This compact representation enables a
direct mapping from the measured output quality to the
shots-to-approximate-solution cost for a target ratio~$r$.

The complete workflow is illustrated in Fig.~\ref{fig:schematic}. An initial atom
array is prepared on a programmable Rydberg platform and a quantum annealing
sweep drives the system toward low-energy independent-set configurations. Each
experimental shot yields a raw bitstring of Rydberg occupations
$z\in\{0,1\}^N$, which is then passed through a deterministic postprocessing consisting of three sequential stages: (i)~\emph{feasibility
projection}, which removes atoms participating in blockade violations;
(ii)~\emph{greedy maximization}, which augments the independent set without
violating constraints; and (iii)~\emph{$\ell$-swap local improvement}, which
accepts local replacements that strictly increase the set size. Closely related
vertex-reduction and vertex-addition procedures were used in the Rydberg MIS
benchmark of Ref.~\cite{Ebadi2022Science}, while related swap-based local
improvement ideas appear in the quantum-enhanced simulated annealing analysis of
Ref.~\cite{Jeong2025QESA}. The final output $\Spost$ is a valid independent set whose distance from the
MIS manifold is quantified by the shell index~$\shell$. Since raw bitstrings
frequently contain blockade violations and thus do not constitute valid
independent sets, all performance analysis---including shell-model fitting,
excitation-matched baseline comparison, and shots-to-approximate-solution
evaluation---operates on $\Spost$ rather than on the raw bitstring. By applying
the same postprocessing to excitation-density-matched random inputs, we isolate
the contribution of the quantum dynamics from trivial density effects.

In the remaining part of the paper, we introduce the $\STS(r)$ framework
and the shell model (Sec.~\ref{sec:framework}), which convert postprocessed
shell statistics into an operational shot cost at an arbitrary target
ratio. Using
this framework together with an excitation-matched random baseline, we isolate
the trivial dependence of postprocessing performance on the raw excitation
density. We then analyze experimental data from programmable Rydberg atom
arrays with up to 125 sites to show that annealing produces a positive residual
shell-concentration advantage and characteristic scaling behavior
(Sec.~\ref{sec:experiment}). We further show that the operational impact of
this advantage depends strongly on the target ratio: it is substantial for
near-exact targets close to the MIS manifold, while becoming small for
sufficiently relaxed approximate targets (Sec.~\ref{sec:discussion}). Finally,
we conclude with a discussion of the two-regime structure and its implications.

\section{Framework}\label{sec:framework}

\subsection{Shots-to-approximate-solution metric} \noindent
For any shot-based hybrid optimizer, we define the
\emph{shots-to-approximate-solution} at target approximation ratio~$r$ as the
number of independent experimental shots required to obtain at least one
postprocessed output satisfying the target ratio~$r$ with confidence~$\preq$:
\begin{equation}
\STS(r;N) \equiv
\left\lceil \frac{\ln(1-\preq)}{\ln(1-p_r)} \right\rceil,
\label{eq:sts_product}
\end{equation}
where $N=|V|$ is the number of vertices, equivalently the number of atoms/sites
in the encoded graph, and $p_r$ is the per-shot success probability at ratio~$r$. Throughout this work, we set $\preq = 0.99$.
In this work, our main focus is on $\STS(r;N)$, since the measured shell
distribution directly determines $p_r$ and thus the shot complexity for a given
target ratio~$r$.

\subsection{Shell representation and effective-\texorpdfstring{$\beta$}{beta}
model}  \noindent
Let $\alpha(G)$ denote the size of the maximum independent set of a graph~$G$.
After deterministic postprocessing, each output is a valid independent set
$\Spost \subseteq V$. We define the shell index as
\begin{equation}
\shell \equiv \alpha(G) - |\Spost|,
\label{eq:shell_def}
\end{equation}
so that $\shell = 0$ labels the optimal (MIS) solutions, while larger values of
$\shell$ correspond to increasingly sub-optimal outputs. For a target
approximation ratio $r$, any output satisfying
$|\Spost| \ge \lceil r\alpha\rceil$ is accepted. This condition defines the
shell cutoff
\begin{equation}
J(r) = \alpha - \lceil r\alpha\rceil ,
\label{eq:J_def}
\end{equation}
with the associated success probability
\begin{equation}
p_r = \sum_{j=0}^{J(r)} \pi_j .
\label{eq:pr_shell}
\end{equation}
Here $\pi_j$ denotes the probability that a single postprocessed shot lies in
shell~$j$, namely that the returned independent set has size $\alpha-j$.

We model the postprocessed shell distribution by the degeneracy-weighted form
\begin{equation}
\pi_j(\beta) =
\frac{d_{\alpha-j}\,e^{-\beta j}}
{\displaystyle\sum_{u\ge 0} d_{\alpha-u}\,e^{-\beta u}},
\label{eq:boltz_shell}
\end{equation}
where $d_{\alpha-j}$ denotes the number of independent sets of size
$\alpha-j$. This form has a natural maximum-entropy interpretation: maximizing
the entropy of $\pi$ relative to the degeneracy measure
$d=\{d_{\alpha-j}\}$ (the negative of the relative entropy),
\begin{equation}
\mathcal{S}[\pi\,\|\,d]
=
-\sum_{j\ge 0}\pi_j
\ln\!\left(\frac{\pi_j}{d_{\alpha-j}}\right),
\end{equation}
subject to normalization and a fixed mean shell depth
$\langle j\rangle_{\pi}=\sum_j j\pi_j$ yields Eq.~\eqref{eq:boltz_shell} as the
unique stationary distribution, with~$\beta$ emerging as the Lagrange
multiplier conjugate to the mean-shell constraint~\cite{Jaynes1957PR}.
Beyond this inference-based motivation, Eq.~\eqref{eq:boltz_shell} also admits
an algorithmic derivation. As shown in Appendix~\ref{app:locality}, the finite
interaction radius of the postprocessing pipeline, applied to input ensembles
with short-range correlations, produces a degeneracy-weighted exponential
shell distribution with corrections of order $\mathcal{O}(j^2/N)$; the shell
model is therefore expected to be most accurate precisely in the low-$j$
region that controls $\STS(r)$ for near-exact targets.
In this work, the primary role of Eq.~\eqref{eq:boltz_shell} is operational. The observed
distributions, as validated in Sec.~\ref{subsec:validation}, are in agreement
with Eq.~\eqref{eq:boltz_shell} using an estimated parameter~$\beta$, which
serves as a single-parameter summary of output quality.

Equation~\eqref{eq:boltz_shell} has the form of a Boltzmann distribution,
and it is natural to ask whether $\beta$ reflects an effective temperature
reached by the annealing dynamics. Analogous questions are extensively studied
for superconducting quantum annealers, whose outputs are commonly modeled as
approximate Boltzmann samples at an instance-dependent effective temperature
set by a freeze-out point of the
dynamics~\cite{Amin2015,Benedetti2016,Marshall2017,Vuffray2022}. What
$\beta$ measures directly is the density of postprocessing-irreparable defects
in the raw output. An effective temperature, if present, would control that
same defect density, so the two readings are naturally related: $\beta$ would
then be a monotone proxy for such a temperature rather than a temperature
itself. The shell distribution alone cannot go further, since it describes the
\emph{postprocessed} output of a many-to-one classical map and, as shown in
Appendix~\ref{app:locality}, the same degeneracy-weighted form arises for any
product-measure input---including the non-thermal Bernoulli baseline.
Distinguishing the possibilities therefore requires the raw bitstring
statistics, a direction we outline in Sec.~\ref{sec:discussion}.

\subsection{Excitation-matched random baseline} \noindent
A direct interpretation of the annealing quality parameter~$\beta_{\mathrm{ann}}$
is complicated by variations in the raw Rydberg excitation density. Let
$\pexc^{(\mathrm{ann})}$ denote the mean fraction of excited atoms per shot in
the annealing data. Because classical postprocessing acts on the raw excitation
pattern, applying it to bitstrings with $N\pexc$ excitations generally yields a
different solution distribution than applying it to bitstrings with
$N\pexc'$ excitations, even when both are sampled uniformly at random. This
effect arises solely from differences in excitation density and is independent
of any structural concentration induced by the annealing dynamics.

To account for this effect, we introduce an excitation-matched random baseline
defined as
\begin{equation}
\betarand\!\left(\pexc^{(\mathrm{ann})}\right)
\equiv
\beta\bigl[
\mathrm{PP}(\mathrm{Bern}(\pexc^{(\mathrm{ann})})^{\otimes N})
\bigr],
\label{eq:beta_rand_def}
\end{equation}
where $\mathrm{Bern}(\pexc^{(\mathrm{ann})})^{\otimes N}$ represents $N$
independent Bernoulli random variables with excitation probability
$\pexc^{(\mathrm{ann})}$, and $\mathrm{PP}(\cdot)$ denotes the deterministic
postprocessing pipeline. The resulting annealing advantage is then given by
\begin{equation}
\Delta\beta_{\mathrm{ann}}
\equiv
\betaann -
\betarand\!\left(\pexc^{(\mathrm{ann})}\right).
\label{eq:delta_beta_def}
\end{equation}
A positive value of $\Delta\beta_{\mathrm{ann}}$ indicates that the annealing
output yields a shell distribution more strongly concentrated toward low-shell
solutions than can be explained by excitation density alone. This Bernoulli
model serves as a minimal density-controlled null model: it matches the mean
excitation density per shot---the dominant one-point statistic---without
attempting to reproduce higher-order spatial correlations present in the
quantum output.

The restriction to a one-point statistic is deliberate. The number of
blockade-violating edges is not an independent knob: its suppression is itself
a signature of the blockade physics under test, so a baseline matched to it
would build part of the structure being tested into the null model, and a
vanishing residual could no longer be interpreted. Matching it \emph{in
addition} to the excitation density is moreover impossible within a product
measure, since a single Bernoulli parameter cannot satisfy two constraints; an
edge-matched product ensemble does exist, but it reproduces a substantially
lower excitation density and therefore no longer controls the one-point
statistic that the construction is designed to control. We therefore use the
density-matched Bernoulli ensemble as the minimal spatially uncorrelated
baseline, and expect an edge-matched variant to shift $\betarand$ only
slightly at the present system sizes.

The locality derivation of Appendix~\ref{app:locality} sharpens the meaning of
this comparison. For any product-measure input, the shell form of
Eq.~\eqref{eq:boltz_shell} follows with an exponent fixed entirely by the
one-point statistics; $\betarand(\pexc)$ therefore represents the quality
parameter attainable by an \emph{arbitrary} spatially uncorrelated input
ensemble at the given excitation density. A positive
$\Delta\beta_{\mathrm{ann}}$ consequently certifies the presence of
correlations in the quantum output, beyond one-point statistics, that suppress
postprocessing-irreparable defects---independently of any assumption about the
microscopic origin of those correlations.

\subsection{Quantum annealing and plateau behavior} \noindent
Under the annealing schedule studied here, the inferred annealing quality
parameter~$\beta_{\mathrm{ann}}(T)$ increases with total sweep time~$T$ and
saturates to a plateau over a finite time window. We denote the plateau value by
$\betaann$. This plateau operationally defines a regime in which the final
postprocessed shell distribution becomes largely insensitive to further
increases in annealing time. The residual quantity
$\Delta\beta_{\mathrm{ann}}=\betaann-\betarand(\pexc^{(\mathrm{ann})})$
then quantifies the concentration advantage of quantum annealing over the
excitation-matched random baseline.

\section{Experimental Results}\label{sec:experiment}

\subsection{Graph instances and experimental setup} \noindent
We have investigated MIS optimization using the programmable Rydberg-atom
Hamiltonian
\begin{equation}
H(t) = \sum_{i=1}^{N} \frac{\Omega(t)}{2}\sigma_i^x
- \sum_{i=1}^{N} \Delta(t)\,n_i
+ \sum_{i<j} U_{ij}\,n_i n_j,
\label{eq:H_rydberg}
\end{equation}
where $i$ and $j$ label atoms in the array, $n_i=(1+\sigma_i^z)/2$ denotes the
local Rydberg occupation number, and $U_{ij}=C_6/r_{ij}^6$ is the van der Waals
interaction~\cite{Singer2005PRL,Saffman2010RMP,BrowaeysLahaye2020NatPhys}. The
graph instances considered here are site-diluted induced subgraphs of the
King's lattice, embedded with lattice spacing $a = 5.0~\mu\mathrm{m}$ and chosen
to satisfy the blockade condition $a < R_b < a\sqrt{2}$~\cite{Pichler2018PNAS,Ebadi2022Science}.

All experiments were performed on QuEra Aquila~\cite{QuEra2026} with $M=500$
shots per instance. The raw measurement records, the graph instances with
their exact maximum-independent-set data, and the analysis code that reproduce
all figures and quoted values are openly available~\cite{Jung2026Data}. The dataset comprises four system-size groups: small
($N=30$--$35$), medium ($N=60$--$65$), large ($N=90$--$95$), and xlarge
($N=120$--$125$), with 30 instances in each group. The exact maximum independent set
sizes, listed in Table~\ref{tab:gap}, average $\bar\alpha=12.2$--$43.9$ across
the four groups and thereby set the granularity $\Delta r \simeq 1/\alpha$ of
the accessible target ratios (Sec.~\ref{sec:discussion}). Within each size
group,
instances are further partitioned into five hardness bins according to
\begin{equation}
H(G)=\frac{d_{\alpha-1}(G)}{\alpha\,d_{\alpha}(G)},
\label{eq:hardness_proxy}
\end{equation}
where $H(G)$ measures the ratio of near-optimal to optimal degeneracy,
with larger values indicating stronger entropic competition from
suboptimal first-shell solutions and thus lower probability of
reaching the MIS~\cite{Ebadi2022Science}. Bin boundaries are placed at equal
intervals between the minimum and maximum $H(G)$ values in the dataset;
instances within each bin span a range of approximately $\pm 0.1$ in $H$.
The lowest-hardness bin (bin~0) is retained in the dataset but excluded
from averaged quantitative analyses, as the postprocessed $\betaann$
values for these instances lie well outside the range of harder bins;
at very low hardness, the postprocessing pipeline effectively resolves
the MIS with high probability, making the shell distribution qualitatively
distinct from the remainder of the dataset.

\subsection{Protocol and excitation-rate measurement}  \noindent
For quantum annealing, the Rabi frequency $\Omega$ is ramped from zero over
$T_{\mathrm{ramp}} = 100~\mathrm{ns}$, then held at
$\Omega = 2\pi \times 1~\mathrm{MHz}$ while the detuning $\Delta$ is swept
linearly from $\Delta_i/2\pi = -3.0~\mathrm{MHz}$ to
$\Delta_f/2\pi = +2.0~\mathrm{MHz}$ over a total evolution time
$T \in \{500,1000,1500,2000,2500,3000,4000,5000\}~\mathrm{ns}$.

For each instance and annealing time, we define the mean excitation density
\begin{equation}
\pexc^{(\mathrm{ann})}(T)
=
\frac{1}{M}\sum_{s=1}^{M}
\frac{1}{N}\sum_{i=1}^{N}\mathbf{1}[z_{s,i}=|r\rangle],
\label{eq:pexc_def}
\end{equation}
where $s$ indexes experimental shots. This quantity is used in constructing the
matched baseline and is therefore defined directly from the raw experimental
data, rather than only after fitting~$\beta$.

\subsection{Excitation-matched baseline construction}  \noindent
The annealing protocol produces $\pexc^{(\mathrm{ann})}\approx 0.26$, varying
across instances. A comparison to a uniform random baseline ($\pexc=0.5$) would
therefore conflate two distinct effects: differences in excitation density and
differences in the structure of the postprocessed outputs. To separate these
contributions, we estimate the excitation-matched random baseline
$\betarand(\pexc^{(\mathrm{ann})})$ using $M=500$ samples of
$\mathrm{Bernoulli}(\pexc^{(\mathrm{ann})})^{\otimes N}$, processed through the
same deterministic postprocessing pipeline. The number of samples is chosen to
match the experimental shot count, ensuring that the statistical uncertainty of
the baseline estimate is comparable to that of the experimentally inferred
$\beta_{\mathrm{ann}}$.

\subsection{Plateau behavior and structural advantage}   
\begin{figure*}[t]
  \centering
  \includegraphics[width=0.95\linewidth]{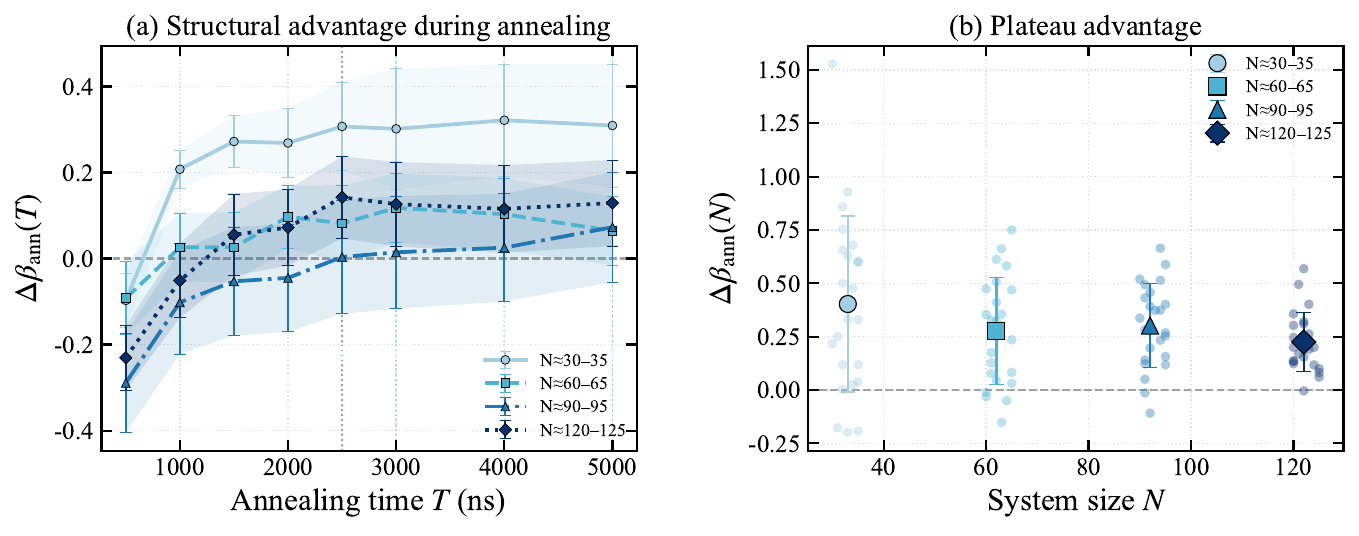}
  \caption{\textbf{Plateau behavior and annealing advantage with
  excitation-matched random baseline.}
  \textbf{(a)}
  $\Delta\betaann(T)
  = \betaann(T) -
  \betarand(\pexc^{(\mathrm{ann})}(T))$ as a function of annealing time for all
  four size groups (bin~2; see Fig.~\ref{fig:anneal_baseline} for the full
  hardness dependence). The random baseline is evaluated at the
  per-instance, per-$T$ excitation rate~$\pexc^{(\mathrm{ann})}(T)$, controlling
  for the density effect at every point of the sweep. In all groups,
  $\Delta\betaann(T)$ rises from negative or near-zero values at
  short times and saturates at a positive plateau near
  $T^{\star} \approx 2500$--$5000~\mathrm{ns}$ (group-dependent).
  \textbf{(b)} Per-instance $\Delta\betaann$ at the plateau
  ($T = 5000~\mathrm{ns}$), for the $24$ instances of bins~1--4 in each size
  group, as a function of system size~$N$. Large symbols are the group means
  of Table~\ref{tab:gap}, with error bars giving the sample standard
  deviation across instances; the advantage is positive in $86$ of the $96$
  instances.}
  \label{fig:anneal_advantage}
\end{figure*}

\noindent
Figure~\ref{fig:anneal_advantage}(a) shows
$\Delta\beta_{\mathrm{ann}}(T)=\beta_{\mathrm{ann}}(T)-
\betarand(\pexc^{(\mathrm{ann})}(T))$ as a function of annealing time~$T$ across
all four size groups. In every group, $\Delta\beta_{\mathrm{ann}}(T)$ increases
from negative or near-zero values at short times and approaches a positive
plateau over a finite time window, empirically observed around
$T^{\star}\approx 2500$--$5000~\mathrm{ns}$ depending on the group. This plateau
quantifies the residual concentration advantage of the annealing dynamics after
accounting for the trivial contribution from excitation density.

\begin{figure*}[t]
  \centering
  \includegraphics[width=0.95\linewidth]{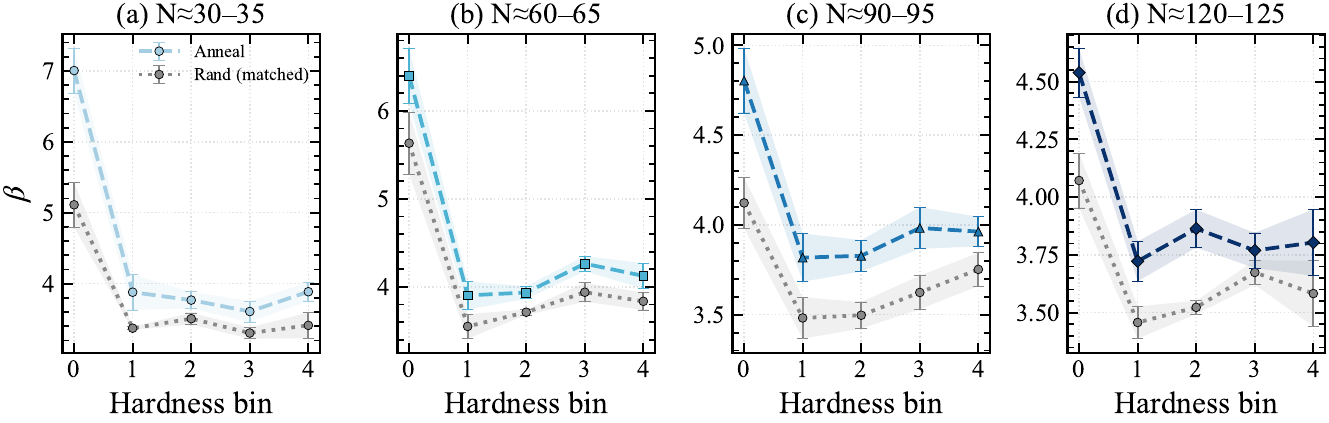}
    \caption{\textbf{Annealing quality parameter and excitation-matched random
      baseline across hardness bins.}
      $\beta_{\mathrm{ann}}$ (blue squares) and the excitation-matched random
      baseline $\betarand(\pexc^{(\mathrm{ann})})$ (gray circles, dashed) as a
      function of hardness bin at $T = 5000~\mathrm{ns}$, shown separately for each
      size group. The excitation
      rate~$\pexc^{(\mathrm{ann})}$ is measured per instance from experimental
      shots, so each random-baseline point is density-matched to the corresponding
      annealing measurement. Annealing consistently exceeds the matched baseline
      across all bins and system sizes, with the gap
      $\Delta\beta_{\mathrm{ann}} = \beta_{\mathrm{ann}} -
      \betarand(\pexc^{(\mathrm{ann})})$ remaining positive throughout. The
      lowest-hardness bin (bin~0) is shown for completeness but is excluded from the
      averaged quantitative summaries because of larger matched-baseline
      fluctuations.}
  \label{fig:anneal_baseline}
\end{figure*}

Figure~\ref{fig:anneal_advantage}(b) shows the per-instance values of
$\Delta\beta_{\mathrm{ann}}$ at the plateau ($T=5000~\mathrm{ns}$) as a
function of system size~$N$. The group means remain positive across all four
size groups.
Two features of Table~\ref{tab:gap} are worth stating explicitly. First,
$\betaann$ is essentially size-independent: the group values span
$3.79$--$4.06$, comparable to the within-group scatter, and a per-instance
regression against~$N$ gives a slope $-5.7\times10^{-4}$ per site
($95\%$~CI $[-2.5,+1.4]\times10^{-3}$, $p=0.57$). This is expected from
Appendix~\ref{app:locality}, where $\beta$ is a log-probability per defect and
hence intensive; the extensive growth of the shot cost enters through
$d_{\alpha-j}$, not~$\beta$. Second, the nominal decrease of
$\Delta\beta_{\mathrm{ann}}$ from $0.40$ (small) to $0.23$ (xlarge) is carried
by the small group alone: over all four groups the slope is
$-1.7\times10^{-3}$ per site ($95\%$~CI $[-3.3,-0.1]\times10^{-3}$, $p=0.04$,
$R^2=0.05$), but excluding small it is consistent with zero
($-8.2\times10^{-4}$, $p=0.40$). The small group is also the only one
departing from the size-collapse of Appendix~\ref{app:degeneracy}, as expected
at $\bar\alpha\approx 12$. We therefore do not read Table~\ref{tab:gap} as
evidence of a systematic decay with system size.
Figure~\ref{fig:anneal_baseline} further shows that $\beta_{\mathrm{ann}}$
consistently exceeds the excitation-matched random baseline across all hardness
bins and system sizes at $T = 5000~\mathrm{ns}$. Table~\ref{tab:gap} summarizes $\betaann$ and
$\Delta\beta_{\mathrm{ann}}$ per size group at $T=5000~\mathrm{ns}$, where the
measured values lie in the plateau regime. The excitation-matched advantage
$\Delta\beta_{\mathrm{ann}}=0.23$--$0.40$ is positive across all groups and
varies only weakly across hardness bins~1--4 compared with the overall
separation between annealing and the matched baseline. The group means in
Table~\ref{tab:gap} understate the consistency of the effect, since the
instance-to-instance scatter is comparable to the mean: at the level of
individual instances, $\Delta\beta_{\mathrm{ann}}>0$ holds for $86$ of the
$96$ instances in bins~1--4 ($90\%$; $21/24$, $20/24$, $22/24$, and $23/24$
from small to xlarge), which is the precise sense in which annealing exceeds
the matched baseline.

\begin{table}[h]
\caption{Measured annealing quality parameter $\betaann$ and excitation-matched
advantage $\Delta\beta_{\mathrm{ann}}$ per size group at
$T = 5000~\mathrm{ns}$ (bins~1--4). The four size groups correspond to small
($N=30$--$35$), medium ($N=60$--$65$), large ($N=90$--$95$), and xlarge
($N=120$--$125$). The mean exact MIS size $\bar\alpha$ per group is listed
for reference; it determines the target granularity $\Delta r \simeq 1/\alpha$
discussed in Sec.~\ref{sec:discussion}. All quoted uncertainties are
sample standard deviations across the $24$ instances of bins~1--4 in each
group.}
\label{tab:gap}
\begin{ruledtabular}
\begin{tabular}{lccc}
Group & $\bar\alpha$ & $\betaann$ & $\Delta\beta_{\mathrm{ann}}$ \\
\hline
small   & $12.2$ & $3.79\pm0.40$ & $0.40\pm0.41$ \\
medium  & $22.5$ & $4.06\pm0.31$ & $0.28\pm0.25$ \\
large   & $33.0$ & $3.90\pm0.26$ & $0.30\pm0.20$ \\
xlarge  & $43.9$ & $3.79\pm0.24$ & $0.23\pm0.14$ \\
\end{tabular}
\end{ruledtabular}
\end{table}

\subsection{Model validation}\label{subsec:validation}
The shell model is validated on two fronts: goodness of fit to the
empirical shell distribution, and monotonic sensitivity of the
inferred~$\beta$ to known degradation in solution quality.

\begin{figure*}[t]
  \centering
  \includegraphics[width=0.95\linewidth]{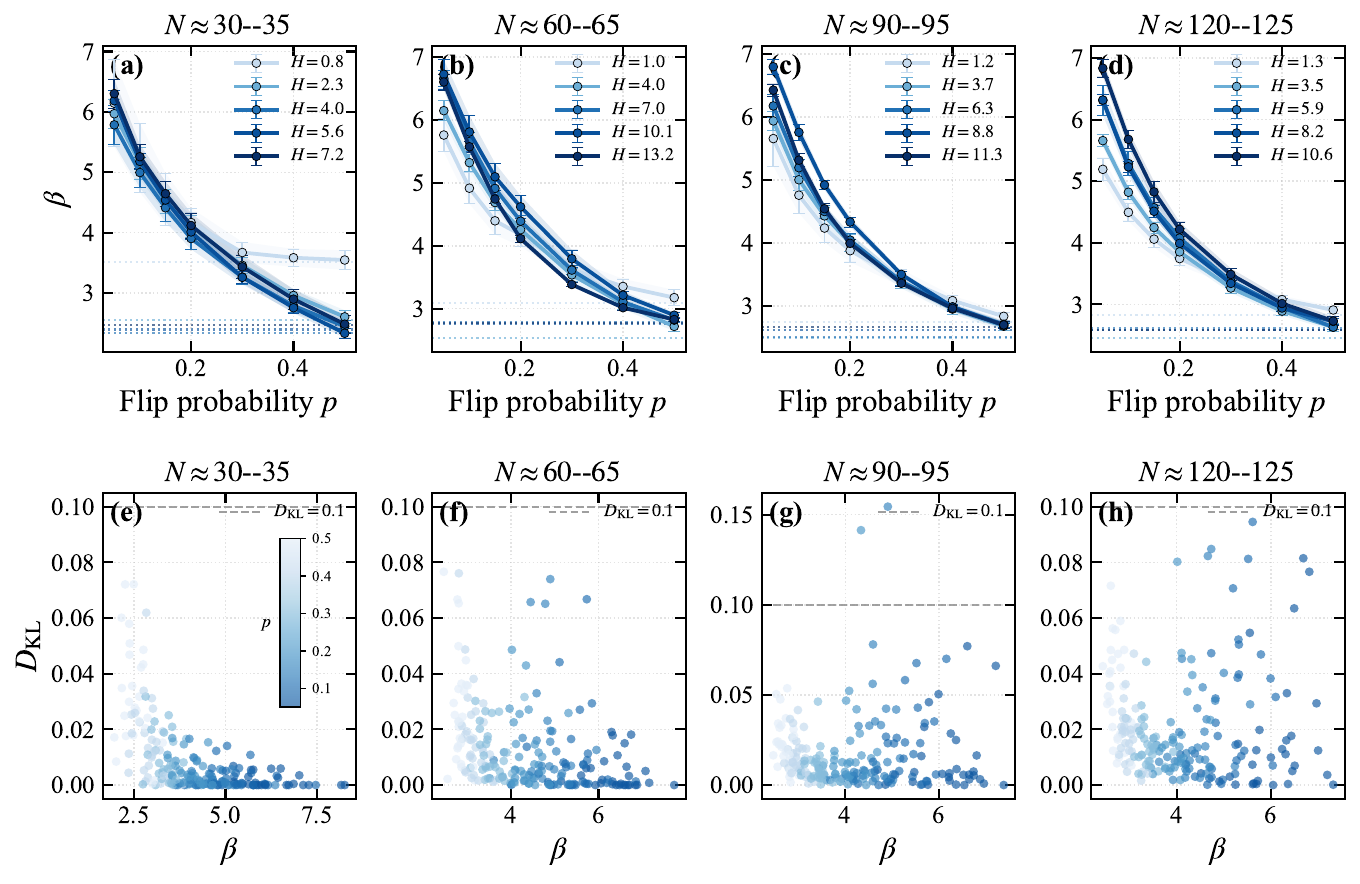}
  \caption{\textbf{Shell-model sensitivity and KL-divergence validation.}
  \textbf{(a)--(d)} Postprocessed~$\beta$ as a function of flip
  probability~$p$ for controlled mock data, constructed by independently
  flipping each bit of the exact MIS solution with probability~$p$ and
  applying the $\ell=1$ postprocessing pipeline, shown separately for each
  size group. Dotted lines indicate the excitation-matched random
  baseline~$\betarand$ for each hardness bin.
  \textbf{(e)--(h)} $D_{\mathrm{KL}}(P_{\mathrm{emp}}\|P_{\mathrm{fit}})$
  between the empirical postprocessed shell distribution and the fitted
  shell model as a function of the inferred~$\beta$, for mock data across
  all four size groups. Color encodes flip probability~$p$; darker points
  correspond to lower~$p$ and hence higher solution quality. The dashed
  line marks $D_{\mathrm{KL}}=0.1$; over $97\%$ of instances fall below
  this threshold.}
  \label{fig:model_validation}
\end{figure*}

To assess sensitivity, we construct controlled mock datasets in which the
distance from the MIS manifold is set by a known flip probability
$p\in\{0.05,0.1,\ldots,0.5\}$. For each graph instance,
$N_{\mathrm{shots}}=2000$ mock measurements are generated by independently
flipping each bit of the exact MIS solution with probability~$p$ and
applying the $\ell=1$ postprocessing pipeline.
Figures~\ref{fig:model_validation}(a)--(d) show that the
inferred~$\beta$ decreases monotonically with~$p$ across all size groups
and hardness bins, confirming that the shell model resolves degradation in
solution quality in the expected direction.

To assess goodness of fit, we use the KL divergence
\begin{equation}
D_{\mathrm{KL}}(P_{\mathrm{emp}} \| P_{\mathrm{fit}})
=
\sum_{j} P_{\mathrm{emp}}(j)
\log\!\frac{P_{\mathrm{emp}}(j)}{P_{\mathrm{fit}}(j)} ,
\label{eq:kl_def_main}
\end{equation}
where $P_{\mathrm{emp}}$ is the empirical postprocessed shell distribution
and $P_{\mathrm{fit}}$ is the corresponding fitted one-parameter model.
Because the magnitude of $D_{\mathrm{KL}}$ depends on the effective
support of the shell distribution, we interpret $D_{\mathrm{KL}}<0.1$ as
an empirical fit-quality threshold within the present dataset rather than
as a universal normalized score.
Figures~\ref{fig:model_validation}(e)--(h) show that over $97\%$ of
tested instances satisfy $D_{\mathrm{KL}}<0.1$, validating the
one-parameter shell description over a broad range of output qualities.
The fitted model is also required to reproduce the dominant shell region
that controls $\STS(r)$; the threshold $D_{\mathrm{KL}}<0.1$ is therefore
used in combination with the monotonic calibration above, so that the
validation checks both goodness of fit and sensitivity of the fitted
parameter.

The mock ensemble also occupies a precise position within the locality
derivation of Appendix~\ref{app:locality}. Since the mock inputs
$z = S^{\star}\oplus\mathrm{Bern}(p)^{\otimes N}$ form an inhomogeneous
product measure, the independence assumptions of the derivation are satisfied
exactly by construction; the exponential shell form and the monotonic decrease
of~$\beta$ with~$p$ then follow as corollaries rather than purely empirical
observations. The derivation further predicts
$\beta_{\mathrm{mock}}(p)\simeq -m\ln p + \mathrm{const}$ as $p\to 0$, where
$m$ is the minimal number of flips required to create a defect that the
$\ell=1$ pipeline cannot repair; the overall decrease of~$\beta$ by
$\approx 3.5$--$4$ across the decade $p=0.05$--$0.5$ in
Figs.~\ref{fig:model_validation}(a)--(d) is consistent with a mixture of
single- and double-flip defect channels. We note an asymmetry in what this
validation establishes: because the mock measure is centered on a single
reference MIS, it samples the shell manifold locally around $S^{\star}$ rather
than degeneracy-uniformly. The mock data therefore calibrate the sensitivity
of the fitted~$\beta$ to known quality degradation, while the goodness-of-fit
test on the experimental data assesses the shell description itself.

Finally, the end quantity can be checked without the model: with $M=500$
shots per instance, the fraction landing in shells $j\le J(r)$ is a
model-free estimate of~$p_r$, available at every admissible cutoff~$J$. At the
exact target the median counts are $146.5$, $45$, $12$, and $5$ per $500$
shots (small to xlarge, annealing, bins~1--4), giving direct-count
$\STS(r=1)=14$, $49$, $190$, and $459$ against per-instance shell-model values
$11$, $27$, $63$, and $140$. The xlarge direct count rests on $5$ exact hits
in $500$ shots and thus carries a $\sim 45\%$ counting uncertainty. The discrepancy is confined to the exact target:
for $J\ge 1$ the two agree to within $25\%$ in the large and xlarge groups,
whereas at $J=0$ the model overestimates $\pi_0$ by up to a factor of
$\approx 3$ ($\approx 1.2$ in the small group, growing to $2$--$3$ in the
larger groups). This is
the shell where the stable-set-count refinement of Sec.~\ref{sec:discussion}
is expected to matter most, since $d_{\alpha}$ counts all maximum independent
sets rather than the $1$-swap-stable ones the pipeline can return. Shell-model
$\STS(r=1)$ should therefore be read as a lower bound at that level. At the
relaxed target the same count gives $\STS(r=0.9)=2$ in every group, so the
two-regime structure holds by direct counting alone, independently of
Eq.~\eqref{eq:boltz_shell}.

\subsection{Shots-to-approximate-solution scaling}\label{sec:scaling} 
\begin{figure*}[t]
  \centering
  \includegraphics[width=\linewidth]{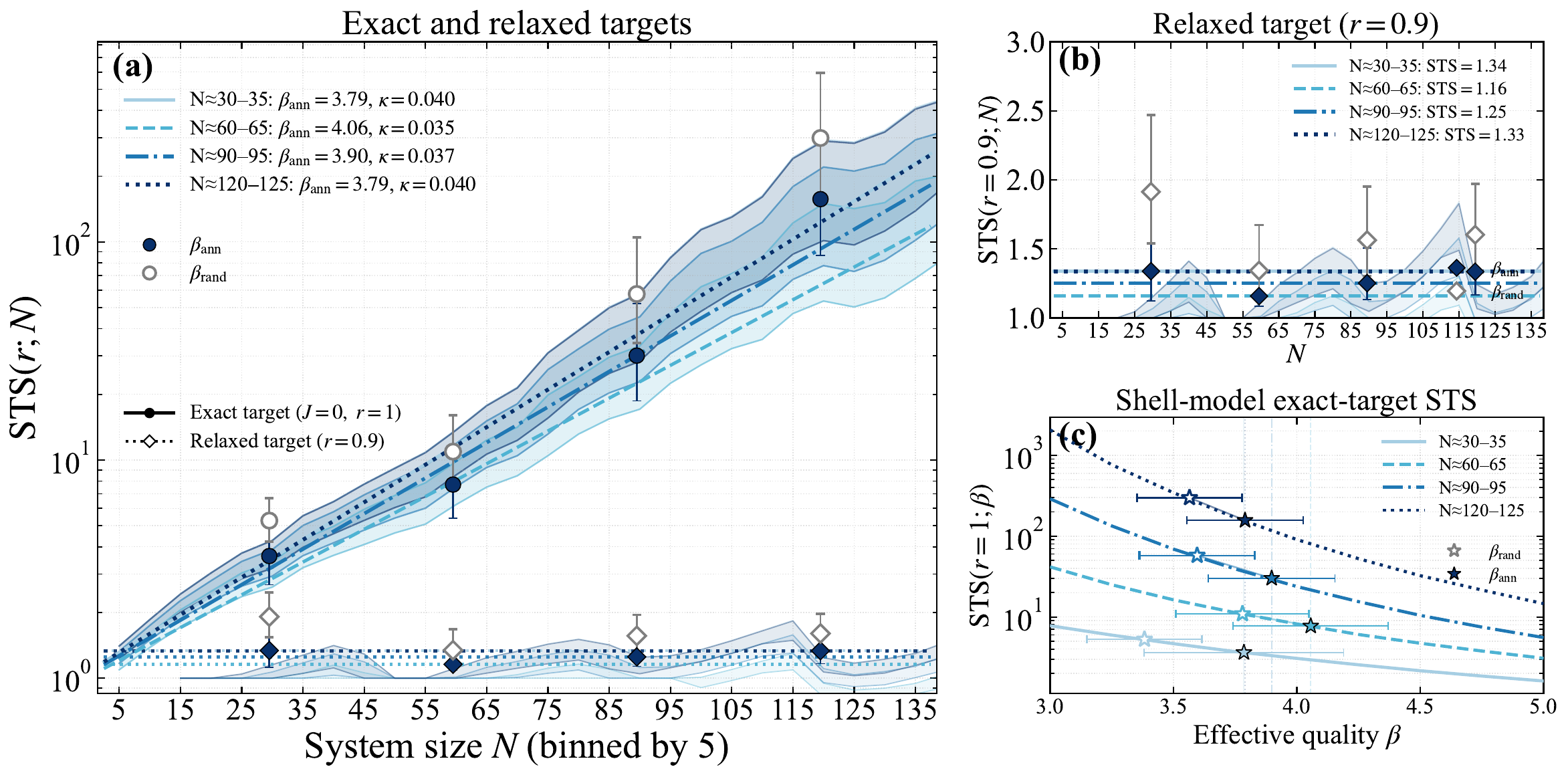}
  \caption{\textbf{Shots-to-approximate-solution scaling.}
  \textbf{(a)} Required shot count $\STS(r=1;N)$ versus system size~$N$ for the
  exact target ($J=0$, equivalently $r=1$). The four curves use the
  group-dependent annealing operating values $\betaann^{(g)}$, so that each
  curve directly corresponds to the experimentally achieved operating point for
  that size group. The curves are exponential fits $\propto e^{\kappa N}$ to the binned
  medians, while
  the shaded regions indicate the interquartile range of the hardest 10\% of
  instances, ranked by $\STS(r=1;N)$, within each size bin. The increase of
  $\STS(r=1;N)$ with~$N$ illustrates the exponential shot cost of reaching the
  exact optimum.
  \textbf{(b)} Expanded view of the relaxed fixed-ratio target ($r=0.9$). The
horizontal guide lines show the shell-model prediction at the same
group-dependent annealing operating values $\betaann^{(g)}$, plotted on a
linear scale because the required number of shots remains close to unity.
Filled diamonds mark $\STS(r=0.9;\betaann^{(g)})$ for the four experimental
size bins, and open diamonds mark the corresponding excitation-matched random
baselines $\STS(r=0.9;\betarand)$.
  \textbf{(c)} Shell-model $\STS(r=1;\beta)$ as a function of effective quality
  for the four experimental size bins. Curves show the median shell-model
  prediction for each size range, filled stars mark the measured annealing
  operating points $\betaann^{(g)}$, and open stars mark the excitation-matched
  random baselines $\betarand$. Faint vertical guide lines indicate the
  annealing operating values $\betaann^{(g)}$ at which the curves in panels~(a,b)
  are evaluated. Together, the panels show that exact optimization retains an
  exponentially growing shot cost, whereas relaxed approximate targets require
  only order-unity shots over the same size range.}
  \label{fig:crossover}
\end{figure*}

\noindent
The fitted shell model provides a direct map from output quality, summarized by
$\beta$, to the required number of shots for a target approximation ratio. We
use this map to compare near-exact and relaxed targets in
Fig.~\ref{fig:crossover}. For each experimental size group~$g$, the guide curves
are evaluated at the measured annealing plateau value~$\betaann^{(g)}$, so that
the exponential fits in panel~(a) directly correspond to the experimentally
achieved operating points.

For the exact target ($J=0$, equivalently $r=1$),
Fig.~\ref{fig:crossover}(a) shows the required shot count $\STS(r=1;N)$ as a
function of system size using these group-dependent values $\betaann^{(g)}$. The
plotted curves are exponential fits to the binned medians, while the shaded
regions indicate the interquartile range of the hardest 10\% of instances,
ranked by $\STS(r=1;N)$, within each size bin. In all cases,
$\STS(r=1;N)$ increases approximately exponentially with $N$, consistent with
the large-deviation interpretation developed in Sec.~\ref{sec:discussion}.
Increasing $\beta$ lowers the required shot count at fixed~$N$, as shown
explicitly in panel~(c); at the present sizes the four fitted growth rates are
nearly indistinguishable ($\kappa=0.035$--$0.040$), so within the shell-model
description the improvement appears chiefly as a prefactor reduction rather
than a resolvable change of slope.

In contrast, Fig.~\ref{fig:crossover}(b) shows an expanded view of the
shell-model prediction for a relaxed fixed-ratio target, $r=0.9$, evaluated at
the same group-dependent plateau values $\betaann^{(g)}$. In this regime, the
required shot count remains close to unity over the entire size range and
changes only weakly across the experimentally relevant range of~$\beta$. The
excitation-matched random baselines, $\STS(r=0.9;\betarand)$, also lie in this
order-unity regime. Thus, although annealing produces a positive structural
shift $\Delta\beta_{\mathrm{ann}}>0$, the operational benefit of this shift is
strongly target-dependent: it is substantial for near-exact targets close to the
MIS manifold, but small for sufficiently relaxed approximate targets.

Figure~\ref{fig:crossover}(c) makes this target dependence explicit from the
opposite perspective. At the exact target, $\STS(r=1;\beta)$ decreases rapidly
with increasing $\beta$, and the separation between size groups grows with
system size. The measured annealing plateau values $\betaann^{(g)}$ lie to the
right of the excitation-matched random baselines $\betarand$, implying fewer
required shots at fixed $N$. The faint vertical lines mark the plateau values
$\betaann^{(g)}$, at which the exponential fits in panel~(a) are evaluated. The
large reduction visible at the exact target does not contradict the near-unity
shot cost in Fig.~\ref{fig:crossover}(b): once the accepted shell range is wide
enough to include the dominant portion of the postprocessed distribution, the
success probability becomes $\mathcal{O}(1)$ and the corresponding shot cost
saturates near one.

Harder instances (larger $H$) correspond to lower postprocessed~$\betaann$, but
the excitation-matched advantage $\Delta\beta_{\mathrm{ann}}$ varies only
weakly across hardness bins within each size group (Table~\ref{tab:gap}). The
main hardness effect is therefore a shift in the overall scale of the shell
distribution, rather than a qualitative change in the
shots-to-approximate-solution scaling structure.

\section{Discussion}\label{sec:discussion}

\subsection{Effective temperature and the meaning of
\texorpdfstring{$\beta$}{beta}} \noindent
A degeneracy-weighted shell distribution characterized by an effective
inverse temperature~$\beta$ describes the postprocessed outputs from
Rydberg quantum annealing experiments. The Boltzmann form invites a
thermal reading of this parameter, familiar from superconducting annealers,
whose outputs are commonly modeled as approximate Boltzmann samples at an
instance-dependent effective
temperature~\cite{Amin2015,Benedetti2016,Marshall2017,Vuffray2022}. Whether
the raw output here admits such an effective-thermal description, and whether
$\betaann$ tracks such a temperature, is a question the shell distribution
cannot decide; it would be settled on the raw bitstring statistics---for
instance the energy distribution of the unprocessed configurations. The two
readings are distinguishable through the plateau: a thermal picture would
have the effective temperature keep falling as the sweep slows, as observed
for quasistatic superconducting annealers~\cite{Amin2015,Marshall2017},
whereas a saturating defect density---the raw-output counterpart of the
plateau---predicts the saturation actually observed in
Fig.~\ref{fig:anneal_advantage}(a).

Across all system sizes, the measured outputs consistently exceed the
excitation-matched random baseline ($\Delta\beta_{\mathrm{ann}} > 0$).
Because this comparison controls for per-shot excitation density, the
residual gap reflects genuine concentration toward low-shell solutions rather
than trivial density effects.

\subsection{Large-deviation origin of the two regimes} \noindent
The measured advantage, $\Delta\beta_{\mathrm{ann}} = 0.23$--$0.40$
(Table~\ref{tab:gap}), acquires direct operational meaning within the
shots-to-approximate-solution framework. In the large-$N$ limit, we introduce
the intensive shell depth $\delta = j/N$. The near-optimal shell degeneracy is
then approximated by an entropy-density form,
\( d_{\alpha-j} \approx \exp[Ns(\delta)] \), where $s(\delta)$ is assumed to be
a concave function over the relevant shell range. The shell weight in
Eq.~\eqref{eq:boltz_shell} therefore behaves as
\begin{equation}
d_{\alpha-j}\,e^{-\beta j} \approx
\exp\!\bigl[N\bigl(s(\delta)-\beta\delta\bigr)\bigr].
\label{eq:ld_weight}
\end{equation}
By the Laplace principle~\cite{DenHollander2000}, the normalization of the shell
distribution is governed by the maximizer
\begin{equation}
\delta_\star
=
\arg\max_{\delta\ge 0}
\left[s(\delta)-\beta\delta\right],
\label{eq:delta_star}
\end{equation}
equivalently satisfying $s'(\delta_\star)=\beta$ when the maximizer lies in the
interior. For a target ratio~$r$, let
$\delta_c(r) = (\alpha - \lceil r\alpha \rceil)/N$ denote the intensive accepted
shell cutoff. The corresponding success probability is controlled by the
difference between the dominant shell weight in the full distribution and the
dominant shell weight within the accepted region:
\begin{equation}
p_r \approx \exp[-N I(\delta_c)],
\label{eq:pr_ld}
\end{equation}
where
\begin{equation}
I(\delta_c)
=
\max_{\delta\ge 0}\left[s(\delta)-\beta\delta\right]
-
\max_{0\le \delta\le \delta_c}
\left[s(\delta)-\beta\delta\right].
\label{eq:rate_function}
\end{equation}
For near-exact targets with $\delta_c<\delta_\star$, the accepted region
excludes the dominant shell, so $I(\delta_c)>0$ and the required shot count
scales as $\STS(r;N)\sim e^{N I(\delta_c)}$. In this regime, increasing~$\beta$
shifts the distribution toward lower shell depths and can reduce the rate
function, allowing even a modest $\Delta\beta_{\mathrm{ann}}$ to translate into
an exponential-in-$N$ reduction in the required shot count within the shell-model
description. By contrast, for sufficiently relaxed targets with
$\delta_c\ge\delta_\star$, the accepted region includes the dominant
contribution to the shell distribution, the rate function vanishes, and
$p_r=\mathcal{O}(1)$, giving $\STS(r;N)=\mathcal{O}(1)$. In this regime, the
same structural shift in~$\beta$ has only a weak operational effect. The
exponential growth of $\STS(r=1;N)$ with $N$ observed at the exact target
($r=1$) and the near-constant shot cost at the relaxed target ($r=0.9$) shown in
Fig.~\ref{fig:crossover}(a,b) are consistent with this two-regime prediction.

\subsection{Finite-size limitations} \noindent
Two features of the present dataset limit how much of this picture is
directly observable, and we state both explicitly.

\begin{figure}[t]
  \centering
  \includegraphics[width=\linewidth]{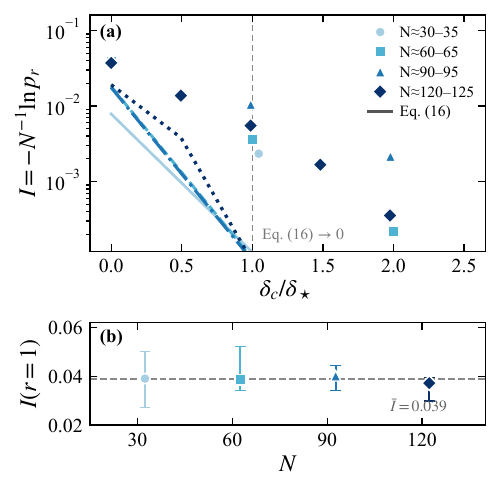}
  \caption{\textbf{Measured rate function across the accessible targets.}
  \textbf{(a)} $I=-N^{-1}\ln p_r$ against $\delta_c/\delta_\star$, obtained by
  counting the same $500$ shots per instance at every admissible cutoff
  $J=0,1,2,\dots$ (median over the $24$ instances of bins~1--4 in each size
  group). Symbols are the direct counts; lines are
  Eq.~\eqref{eq:rate_function} evaluated on the exact degeneracies at the
  per-instance fitted~$\beta$, which vanishes identically at
  $\delta_c=\delta_\star$ (vertical dashed line), the lines dropping below
  the logarithmic scale there. The measured $I$ instead decays smoothly and passes through
  $\delta_\star$ without a break: the kink predicted in the $N\to\infty$
  limit is not yet present.
  \textbf{(b)} $I$ at the exact target versus~$N$ (median and interquartile
  range across instances): the group medians agree to within $8\%$ (dashed
  line, $\bar I=0.039$), well inside the instance-to-instance scatter, so the
  rate function is intensive and $NI$ grows linearly with~$N$. The present
  values, $NI=1.3$--$4.5$, are still of order unity, which is why the Laplace
  step underlying the kink does not yet apply.}
  \label{fig:rate_function}
\end{figure}

First, the cutoff $J(r)=\alpha-\lceil r\alpha\rceil$ is integer-valued, so
targets are quantized in steps of $\Delta r \simeq 1/\alpha$---from
$\approx 0.082$ (small) to $\approx 0.023$ (xlarge). The continuous crossover
at $\delta_c=\delta_\star$ implied by
Eqs.~\eqref{eq:delta_star}--\eqref{eq:rate_function} is therefore not
resolvable here, the two targets being separated by only a handful of
admissible cutoffs. \emph{Those cutoffs are nonetheless measurable.} Counting
the same shots at $J=0,1,2,\dots$ yields $p_r$, and hence the rate function
$I=-N^{-1}\ln p_r$, at every accessible target;
Fig.~\ref{fig:rate_function}(a) shows the result. Two features are relevant
here. (i)~$I$ decays smoothly and approximately geometrically with~$J$ and
passes through $\delta_c=\delta_\star$ without a break, whereas
Eq.~\eqref{eq:rate_function} vanishes identically above $\delta_\star$: the
kink is an $N\to\infty$ feature and is absent at these sizes. This is
expected, since the Laplace step requires $NI\gg 1$ while the measured values
are $NI=1.3$, $2.4$, $3.7$, and $4.5$ at the exact target. (ii)~$I$ at the exact target is
nonetheless independent of system size to within $8\%$
[Fig.~\ref{fig:rate_function}(b)], i.e.\ the rate function is intensive, so
$NI$ grows linearly with~$N$. The sharpening is thus not an extrapolation but
a measured trend, and it places the onset of a resolvable kink
($NI\gtrsim 10$) near $N\approx 260$, roughly twice the size of the present arrays. A fixed ratio is also not a fixed physical relaxation:
$r=0.9$ admits a one-atom deficit in the small group but four in xlarge, so
the uniformity of Fig.~\ref{fig:crossover}(b) conceals a target that weakens
with system size; reporting $\STS$ at fixed $\delta_c$ would remove the
ambiguity but is indistinguishable at the present~$\alpha$. The picture should
sharpen as $\alpha$ grows: $\delta_c(r)$ becomes quasi-continuous and
$I(\delta_c)$ develops a non-analytic point at $\delta_c=\delta_\star$,
vanishing above it and rising with slope $-\beta+s'(\delta_c)$ below. At
finite $\alpha$ the transition is smeared, as
Fig.~\ref{fig:rate_function}(a) shows directly. Confirming the kink itself
requires larger $\alpha$; extrapolating from Eq.~\eqref{eq:boltz_shell} alone
would describe the shell model, not the hardware.

Second, the asymptotic behavior of $\Delta\beta_{\mathrm{ann}}$
remains open at the present sizes. As shown in Sec.~\ref{sec:experiment}, its
apparent decrease is driven by the small group, which also exhibits the
strongest finite-size effects in Appendix~\ref{app:degeneracy}. The remaining
three groups show no resolvable size dependence over $N=60$--$125$. Whether
the advantage saturates or slowly decays therefore remains open and requires
larger systems.

This interpretation relies on two assumptions: that shell counts admit an
entropy-density description at large~$N$, and that the Laplace principle governs
the shell sum. Under these conditions, the fitted parameter~$\beta$ provides a
direct link between experimentally measured output distributions and the scaling
of shot complexity. The exact rescaled counts of Appendix~\ref{app:degeneracy} support
this description over the accessible window, though the window remains too
narrow for a definitive verification. The resulting two-regime structure---exponential sensitivity
for near-exact targets and weak dependence for relaxed ones---is consistent with
analogous observations in gate-based optimization~\cite{Larkin2022QST,Chernyavskiy2025}
and demonstrates that similar behavior arises in analog Rydberg systems when
excitation-density effects are properly controlled.

\subsection{Classical reference cost}\label{subsec:classical} \noindent
The instances studied here are site-diluted King's-lattice subgraphs, a
unit-disk family that admits polynomial-time approximation
schemes~\cite{Hunt1998,Nieberg2005,Erlebach2005} and that is tractable for
exact classical solvers at the sizes accessible on current hardware. The
$\STS(r)$ values reported above therefore characterize the neutral-atom
protocol; they are not evidence of an advantage over classical methods. To make
this explicit and to allow the measured shot counts to be read in absolute
terms, we state the classical reference cost at both ends of the target range.

\emph{Exact target.} The degeneracy computation of
Appendix~\ref{app:degeneracy}---a geometry-aware transfer-matrix dynamic
program over the rows of the King's lattice, run independently on each
connected component---returns the exact maximum independent set size
$\alpha(G)$ together with the near-optimal counts $d_{\alpha-j}(G)$, and thus
constitutes an exact classical solve of every instance in the dataset. Its
cost is reported in Appendix~\ref{app:classical_cost}:
Fig.~\ref{fig:sm_classical}(c) shows the runtime versus~$N$, and
Fig.~\ref{fig:sm_classical}(a) plots the measured $\STS(r{=}1)$ directly
against it. On one core of an Apple~M3
under CPython~3.9.6 the median time per instance rises from $0.6~\mathrm{ms}$
at $N=30$--$35$ to $97~\mathrm{ms}$ at $N=120$--$125$, and all $120$ instances
are solved exactly in under four seconds. Exact optimization on this family is
not polynomial---the PTAS of Refs.~\cite{Hunt1998,Nieberg2005,Erlebach2005}
concerns approximation---but it is subexponential, $2^{\Theta(\sqrt{n})}$, and
this is optimal under the exponential-time
hypothesis~\cite{deBerg2020}. The dynamic program has the corresponding
state-count scaling: its cost is set by the number of transfer-matrix states
per lattice row, bounded by $\varphi^{L}$ with $\varphi$ the golden ratio and
$L=\Theta(\sqrt{N})$ the lattice width. The measured times track this state
count closely ($R^2=0.9996$ against rows${}\times{}$states$^2$) and follow the
guide $\propto e^{0.91\sqrt{N}}$ of Fig.~\ref{fig:sm_classical}(c) out to $N\approx 200$, using
synthetic instances at the same site density beyond the experimental range.
The implementation is unoptimized pure Python, so the constant factor---but
not the scaling---bounds the classical cost from above. The shell-model median
$\STS(r=1;N\!\approx\!120)=157$ shots (binned-fit; the per-instance
shell-model median is $140$)---or the larger direct count of
Sec.~\ref{subsec:validation}---should be read against them: the classical
solve is inexpensive here, and the shot count measures how a protocol
concentrates probability near the MIS manifold, not a competitive cost.

\emph{Relaxed target.} The excitation-matched baseline is a randomized
greedy MIS heuristic with $1$-swap local search
(Algorithm~\ref{alg:postprocessing}), so its shot cost directly provides a
classical reference. The open diamonds of Fig.~\ref{fig:crossover}(b) give
$\STS(r=0.9;\betarand)=1.9$, $1.3$, $1.6$, and $1.6$ from small to xlarge
(medians of the unrounded expression in Eq.~\eqref{eq:sts_product},
which the ceiling turns into two shots in every group): one to two greedy
passes reach the target throughout. A pass costs
$0.05$--$1.6~\mathrm{ms}$ per instance over $N=30$--$200$, growing
polynomially [$\propto N^{1.9}$; Fig.~\ref{fig:sm_classical}(b,c)]. The allowed deficit
$J(0.9)$ takes the constant values $1$, $2$, $3$, and $4$ in the four groups,
so the target is a genuinely multi-atom relaxation only for the largest
instances; the largest baseline cost ($1.9$) occurs in the small group, where
$J(0.9)=1$ places it a single shell from exact. The classical cost is thus
milliseconds throughout, and this relaxed target has limited discriminatory
power as a benchmark for this graph family.

\subsection{Outlook} \noindent
Our study warrants several directions for further investigation. The
large-deviation framework underlying the scaling argument should be tested
through systematic finite-size analysis, including numerical validation of the
entropy-density description and the stability of the inferred exponent across
instances. Strictly, the outputs of Algorithm~\ref{alg:postprocessing} are
$1$-swap-stable maximal independent sets, so the degeneracy entering
Eq.~\eqref{eq:boltz_shell} is in principle the count of such stable sets rather
than of all independent sets of a given size; within the present analysis this
distinction is absorbed into the fitted~$\beta$, and replacing $d_{\alpha-j}$
by stable-set counts is a natural refinement of the degeneracy computation in
Appendix~\ref{app:degeneracy}. A deeper physical interpretation of the
effective parameter~$\beta$ is also needed, particularly its connection to
microscopic features of the annealing dynamics such as diabatic excitations or
spectral structure. Within the locality picture of
Appendix~\ref{app:locality}, $\betaann(T)$ is set by the density of
postprocessing-irreparable defects in the raw output, so quench theories such
as Kibble--Zurek scaling~\cite{Kibble1976,Zurek1985,Keesling2019} enter
through the raw-bitstring correlation length
$\xi(T)$ filtered by the finite repair radius of the pipeline; the natural
experimental probe is therefore the two-point correlation length of the raw
bitstrings---and the coincidence of its saturation with the plateau time
$\Tstar$---rather than the postprocessed shell statistics themselves, which
the pipeline drives toward the form of Eq.~\eqref{eq:boltz_shell} largely
independently of the input correlations. The excitation-matched random
baseline could be refined to incorporate higher-order correlations in the raw
bitstrings, enabling a more complete separation of structural and
density-driven effects. From a benchmarking perspective, extending the
analysis to other graph families and incorporating explicit comparisons with
classical algorithms would clarify the generality and practical significance
of the observed structural advantage. We do not claim an end-to-end quantum
speedup over optimized classical algorithms; rather, the goal is to provide a
density-controlled operational diagnostic of how a given neutral-atom protocol
concentrates probability mass near the MIS manifold and how that concentration
translates into approximation-dependent shot cost. Finally, a systematic
characterization of the plateau time $\Tstar(N)$ and associated hardware
overheads will be needed to connect the present shot-count analysis to
end-to-end runtime performance.

\section{Conclusion}\label{sec:conclusion} \noindent
We have developed a shots-to-approximate-solution framework for evaluating
quantum optimization on neutral-atom platforms, centered on the metric
$\STS(r)$. Experimentally, postprocessed outputs are well captured by a
degeneracy-weighted shell model, enabling a compact and quantitative
characterization of solution quality through the effective parameter~$\beta$.
By introducing an excitation-matched random baseline, we isolate a genuine
structural advantage of quantum annealing, $\Delta\beta_{\mathrm{ann}} > 0$,
across all studied system sizes. This advantage has clear operational
consequences: it yields an exponential reduction in required shot count for
sufficiently near-exact targets within the shell-model description, while
becoming negligible for relaxed targets, where the classical component of
the pipeline alone reaches the target in order-unity attempts. Our study thus
establishes both a practical methodology for benchmarking quantum optimization
and a clear criterion for when quantum protocols provide practical benefits.

\begin{acknowledgments} \noindent
The authors thank J.~Park and M.~P.~Soegianto for fruitful discussions. This
work was supported by the National Research Foundation of Korea (NRF) under
Grant Nos.~RS-2024-00340652 and RS-2025-25464441, and by the Institute of
Information \& Communications Technology Planning \& Evaluation (IITP) under
Grant No.~IITP-2025-RS-2024-004371919, funded by the Korea government (MSIT).
\end{acknowledgments}

\appendix
\section{Degeneracy computation}\label{app:degeneracy}

Near-optimal counts $d_{\alpha-j}(G)$ for
$j=0,\dots,J_{\max}$ are computed by geometry-aware dynamic programming
exploiting the row-by-row structure of King's-lattice adjacency. For larger
instances, we decompose the graph into connected components $\{G_c\}$, compute
size-count polynomials
\begin{equation}
P_c(x)=\sum_m d_m(G_c)x^m,
\end{equation}
and combine them through convolution,
\begin{equation}
\sum_m d_m(G)x^m = \prod_c P_c(x).
\end{equation}
After each convolution step, only coefficients that can still contribute to the
final window $m\in[\alpha-J_{\max},\alpha]$ are retained. This yields the exact
near-optimal shell degeneracies required by Eq.~\eqref{eq:boltz_shell}.

Because the procedure returns $\alpha(G)$ exactly, it also serves as the
exact classical solver of Sec.~\ref{subsec:classical}. We use $J_{\max}=15$
throughout. The median single-core times per instance are $0.6$, $4.4$, $26$,
and $97~\mathrm{ms}$ from small to xlarge, and the full dataset of $120$
instances is processed in $3.95~\mathrm{s}$; the extension to synthetic
instances with $N\approx 200$ is reported in Fig.~\ref{fig:sm_classical}(c)
(Appendix~\ref{app:classical_cost}).

\begin{figure}[t]
  \centering
  \includegraphics[width=\linewidth]{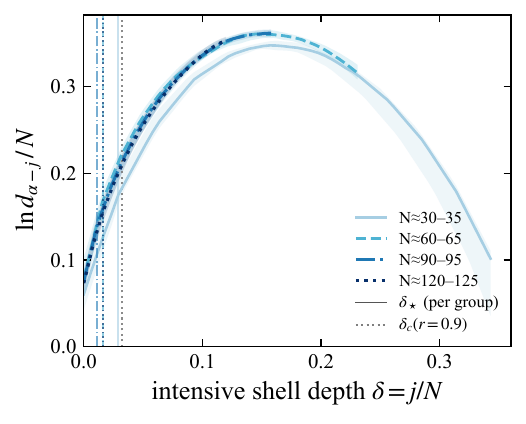}
  \caption{\textbf{Exact near-optimal degeneracy across system sizes.}
  Rescaled shell entropy $\ln d_{\alpha-j}/N$ as a function of the intensive
  shell depth $\delta = j/N$, computed exactly by the dynamic-programming
  procedure of this appendix. Curves show the median over the $24$ instances
  of bins~1--4 in each size group; shaded bands give the interquartile range.
  Vertical lines mark $\delta_\star$ from Eq.~\eqref{eq:delta_star} evaluated
  at the measured plateau values $\betaann^{(g)}$; the gray dotted line marks
  the accepted cutoff $\delta_c(r{=}0.9)$, which lies above $\delta_\star$ in
  all groups. The computed window $j \le J_{\max}=15$ corresponds to a
  group-dependent range $\delta \lesssim J_{\max}/N$; the curves extend to
  $\delta\approx 0.34$, $0.23$, $0.16$, and $0.12$ from small to xlarge, and
  the collapse statement refers to the common window $\delta\lesssim 0.12$.
  The curves are concave over their full ranges; the medium, large, and xlarge
  groups collapse onto a common curve, while the small group lies
  systematically below, consistent with boundary effects at
  $\bar\alpha\approx 12$. The window is too narrow to verify the
  asymptotic entropy-density form assumed in Sec.~\ref{sec:discussion}.}
  \label{fig:degeneracy}
\end{figure}

Figure~\ref{fig:degeneracy} compares the exact rescaled shell counts
across size groups. The finite-size deviations diminish systematically with
increasing~$N$: already from $N\approx 60$ the medium, large, and xlarge
curves collapse onto a common envelope over the shared window
$\delta\lesssim0.12$, and only the smallest group ($\bar\alpha\approx 12$)
lies visibly below it, so the data move toward an $N$-independent
entropy-density form as the system grows. The curves are concave over the
computed range, though the accessible window remains too narrow for a
definitive verification of the asymptotic form. In all groups,
$\delta_\star<\delta_c(r=0.9)$, consistent with the two-regime
interpretation.

\section{Deterministic postprocessing pipeline}\label{app:postproc}

Raw bitstrings $z\in\{0,1\}^N$ are mapped to valid independent sets~$\Spost$
through three sequential stages: (i)~\emph{feasibility projection}, in which
vertices participating in violated edges are removed iteratively (higher-degree
endpoint first); (ii)~\emph{greedy maximization}, in which vertices are added
in ascending order whenever none of their neighbors is already present;
(iii)~\emph{$\ell$-swap improvement}, in which local replacements involving up
to~$\ell$ occupied neighbors are accepted if they strictly increase the set size.
The first two stages are closely related to vertex-reduction and vertex-addition
postprocessing used in prior Rydberg MIS experiments~\cite{Ebadi2022Science},
and related local-improvement ideas appear in recent Rydberg hybrid optimization
work~\cite{Jeong2025QESA}. The procedure repeats until no improving move remains.

\begin{algorithm}[H]
\caption{Deterministic $\ell$-swap postprocessing}
\label{alg:postprocessing}
\begin{algorithmic}[1]
\STATE \textbf{Input:} Graph $G=(V,E)$, bitstring $z\in\{0,1\}^{|V|}$, depth $\ell\ge 0$
\STATE \textbf{Output:} Valid maximal independent set $\Spost$
\STATE $S \leftarrow \{v\in V \mid z_v=1\}$
\STATE \textit{Phase 1: Feasibility projection}
\WHILE{$\exists (u,v)\in E$ such that $u,v\in S$}
  \STATE Remove the endpoint with larger graph degree; break ties by index
\ENDWHILE
\STATE \textit{Phase 2: Greedy maximization}
\FOR{each $v\in V\setminus S$ in ascending order}
  \IF{$N(v)\cap S=\emptyset$}
    \STATE $S \leftarrow S\cup \{v\}$
  \ENDIF
\ENDFOR
\STATE \textit{Phase 3: $\ell$-swap improvement}
\REPEAT
  \STATE $\mathrm{improved}\leftarrow \mathrm{false}$
  \FOR{each $v\in V\setminus S$ in ascending order}
    \STATE $B_v \leftarrow N(v)\cap S$
    \IF{$0<|B_v|\le \ell$}
      \STATE $\hat{S} \leftarrow (S\setminus B_v)\cup\{v\}$; apply Phase~2 to $\hat{S}$
      \IF{$|\hat{S}|>|S|$}
        \STATE $S \leftarrow \hat{S}$
        \STATE $\mathrm{improved}\leftarrow \mathrm{true}$; \textbf{break}
      \ENDIF
    \ENDIF
  \ENDFOR
\UNTIL{$\neg\,\mathrm{improved}$}
\STATE \textbf{return} $S$
\end{algorithmic}
\end{algorithm}

\section{Shell-model form from algorithmic locality}\label{app:locality}

This appendix shows that the degeneracy-weighted exponential form of
Eq.~\eqref{eq:boltz_shell} follows from structural properties of the
postprocessing pipeline applied to input ensembles with short-range
correlations, complementing the maximum-entropy motivation of
Sec.~\ref{sec:framework}. The derivation makes explicit the assumptions under
which Eq.~\eqref{eq:boltz_shell} holds as an asymptotic statement and locates
the leading corrections.

\subsection{Pushforward identity}

Let the input bitstring $z\in\{0,1\}^N$ be drawn from a product measure
$\mu$ (independent coordinates), and let $S=\mathrm{PP}(z)$ denote the output
of Algorithm~\ref{alg:postprocessing}. Exactly,
\begin{equation}
\pi_j \;=\; \sum_{S:\,|S|=\alpha-j} W(S),
\qquad
W(S) \;=\; \Pr_{z\sim\mu}\!\left[\mathrm{PP}(z)=S\right].
\label{eq:pushforward}
\end{equation}
For the homogeneous Bernoulli measure $\mu=\mathrm{Bern}(p)^{\otimes N}$,
\begin{equation}
W(S) = (1-p)^N\, G_S\!\left(\frac{p}{1-p}\right),
\qquad
G_S(x) = \!\!\sum_{z\in\mathrm{PP}^{-1}(S)}\!\! x^{|z|},
\label{eq:basin_enum}
\end{equation}
where $G_S$ is the weight enumerator of the basin of attraction of~$S$ under
the deterministic pipeline. Equation~\eqref{eq:boltz_shell} emerges when
$W(S)$ depends on~$S$ only through its shell index~$j$ (within-shell
exchangeability) and does so exponentially,
$\ln W \approx \mathrm{const} - \beta j$ (additivity). We justify the two
properties in turn.

\subsection{Defect factorization}

The pipeline is local: each of the three phases makes accept/reject decisions
based on graph neighborhoods of bounded radius, so there is an effective
propagation radius $R=\mathcal{O}(1)$ such that whether a given local defect
survives postprocessing is determined by the input restricted to its
distance-$R$ neighborhood [assumption~(i)]. The iterated Phase-3 loop could in
principle propagate information over longer distances through cascades of
accepted swaps; for $\ell=1$ on blockade graphs the accepted moves are
strictly size-increasing and cascades terminate after few steps, and we treat
the finite radius as an assumption whose adequacy is reflected in the fit
quality of Sec.~\ref{subsec:validation}.

An output in shell~$j$ differs from a maximum independent set by~$j$ unit
deficits, which for $1$-swap-stable outputs are localized: whether a deficit
at a given location persists is a function of the input within distance~$R$ of
that location. Because the input measure is a product measure, events
depending on disjoint coordinate sets are exactly independent. Hence, whenever
the defects supporting~$S$ are mutually separated by more than~$2R$
[assumption~(ii), defect sparsity],
\begin{equation}
W(S) \;=\; P_{\mathrm{bulk}}(p)\,\prod_{k=1}^{j} c_k(p),
\label{eq:defect_factorization}
\end{equation}
where $P_{\mathrm{bulk}}$ is the probability that the remainder of the array
is resolved without defects and $c_k(p)$ is the probability of the local input
pattern that creates---and prevents the repair of---defect~$k$. If all defects
are of a single dominant local type [assumption~(iii), type uniformity],
$c_k=c(p)$ and
\begin{equation}
W(S) \;\approx\; P_{\mathrm{bulk}}(p)\, e^{-\beta j},
\qquad
\beta(p) = -\ln c(p).
\label{eq:additivity}
\end{equation}
The status of the three assumptions differs. Assumption~(i) is a property of
the algorithm, subject to the cascade caveat above. Assumption~(ii) holds
asymptotically at low~$j$: for typical defect placements, the probability that
two defects fall within interaction range scales as $\mathcal{O}(j^2/N)$.
Assumption~(iii) is a genuine approximation, but relaxing it merely replaces
$c(p)$ by a type-averaged effective value while preserving the exponential
dependence on~$j$, since the log-cost remains additive over defects. The
exponential form is therefore robust, with the value of~$\beta$ renormalized
by the defect-type mixture; a broader defect-type mixture flattens the
effective exponent, consistent with the lower fitted~$\beta$ observed for
harder instances.

\subsection{Within-shell exchangeability}

The bulk factor $P_{\mathrm{bulk}}$ in Eq.~\eqref{eq:defect_factorization}
depends on the graph but not on the placement of the defects, and on
homogeneous regions of the King's lattice the local weights $c(p)$ are
position-independent. Two residual sources of position dependence remain:
boundary effects, which involve a vanishing fraction of sites at large~$N$,
and the deterministic ascending vertex order used in Phases~2--3, which weakly
biases which of several equivalent outputs is returned. The latter
redistributes weight among outputs within the same shell and is averaged over
by the shell sum in Eq.~\eqref{eq:pushforward}; randomizing the vertex order
per shot would restore within-shell exchangeability exactly in expectation.
Granting exchangeability, the shell sum gives
$\pi_j \propto d_{\alpha-j}\, e^{-\beta j}$.

\subsection{Exactly solvable case and corrections}

The structure above is exact when the blockade graph decomposes into disjoint
clusters, each contributing one MIS slot. The pipeline then factorizes over
clusters, each slot is resolved independently with some probability $q(p)$,
and
\begin{equation}
\pi_j = \binom{\alpha}{j}(1-q)^j q^{\alpha-j}
\;\propto\; d_{\alpha-j}\, e^{-\beta j},
\qquad
\beta = \ln\frac{q}{1-q},
\label{eq:anchor}
\end{equation}
with $d_{\alpha-j}=\binom{\alpha}{j}$ the exact shell degeneracy. The
King's-lattice instances correspond to coupling such clusters with
finite-range interactions; Eq.~\eqref{eq:anchor} is then the zeroth order of
an expansion in defect adjacencies, whose leading correction involves pairs of
interacting defects:
\begin{equation}
\pi_j \;=\; d_{\alpha-j}\, e^{-\beta(p)\, j}
\left[1+\mathcal{O}\!\left(j^2/N\right)\right].
\label{eq:corrected_shell}
\end{equation}

\subsection{Consequences}

Three consequences are used in the main text. First, for any product-measure
input, $\beta$ is fixed entirely by the one-point statistics through $c(p)$;
the excitation-matched baseline $\betarand(\pexc)$ therefore represents the
quality parameter attainable by an arbitrary spatially uncorrelated ensemble
at the given excitation density, and $\Delta\beta_{\mathrm{ann}}>0$ certifies
correlations in the quantum output, beyond one-point statistics, that suppress
postprocessing-irreparable defects (Sec.~\ref{sec:framework}). Second, the
correction in Eq.~\eqref{eq:corrected_shell} is smallest at low~$j$, which is
precisely the region that controls $\STS(r)$ for near-exact targets,
consistent with the fit quality reported in Sec.~\ref{subsec:validation}.
Third, the mock ensemble of Sec.~\ref{subsec:validation},
$z=S^{\star}\oplus\mathrm{Bern}(p)^{\otimes N}$, is an inhomogeneous product
measure, so the independence used in Eq.~\eqref{eq:defect_factorization} holds
exactly by construction; expanding $c(p)=a_m p^m+\mathcal{O}(p^{m+1})$, where
$m$ is the minimal number of flips creating an irreparable defect, predicts
$\beta_{\mathrm{mock}}(p)\simeq -m\ln p+\mathrm{const}$ as $p\to 0$. Strictly,
the degeneracy weighting appropriate to
Algorithm~\ref{alg:postprocessing} counts $1$-swap-stable independent sets of
each size; this distinction, together with the locally sampled
($S^{\star}$-centered) character of the mock measure, is absorbed into the
fitted~$\beta$ and discussed in Secs.~\ref{subsec:validation}
and~\ref{sec:discussion}.

\section{Classical reference cost versus shot cost}
\label{app:classical_cost}

\begin{figure*}[t]
  \centering
  \includegraphics[width=0.98\linewidth]{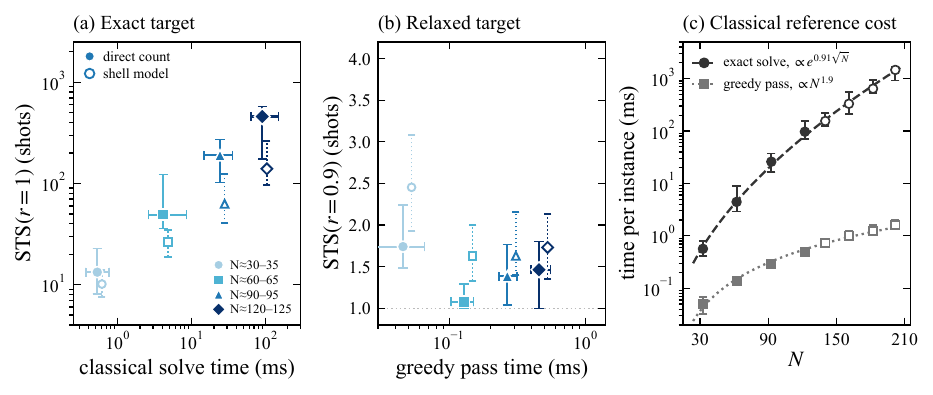}
  \caption{\textbf{Shot cost versus classical reference cost.}
  \textbf{(a)}~Exact target ($r=1$): measured $\STS(r{=}1)$ against the
  median single-core runtime of the exact transfer-matrix solver of
  Appendix~\ref{app:degeneracy}, one point per size group. Filled symbols are direct shot counts; open symbols are the
  per-instance shell model. \textbf{(b)}~Relaxed target: the same comparison at the group cutoff
  $J(0.9)$, against the cost of a single greedy pass on a Bernoulli seed at
  the measured excitation density; the dotted line marks one shot.
  \textbf{(c)}~The classical reference cost versus~$N$ for both cases: exact
  solver (circles, dashed guide $\propto e^{0.91\sqrt{N}}$, the scaling that
  is optimal under the exponential-time hypothesis for unit-disk-type
  graphs~\cite{deBerg2020}) and one greedy pass (squares, dotted guide
  $\propto N^{1.9}$). Whiskers span the full range within each size bin;
  open symbols are synthetic instances extending the range to
  $N\approx 200$. In panels (a) and (b), vertical bars span the
  interquartile range across the $24$ instances of bins~1--4, and horizontal
  bars span the full range of per-instance times; shell-model error bars are
  dotted, and the paired points are slightly offset horizontally for
  visibility.}
  \label{fig:sm_classical}
\end{figure*}

This appendix collects the measurements behind the classical reference
cost of Sec.~\ref{subsec:classical} and compares them directly with the
measured shot cost. All timings are single-core wall-clock times per instance
(one core of an Apple~M3 under CPython~3.9.6); the implementation is
unoptimized pure Python, so the constant factor---but not the
scaling---bounds the classical cost from above. Beyond the experimental
range, the timings are extended to $N\approx 200$ using synthetic instances:
site-diluted King's lattices drawn at the experimental site density
($\approx 0.63$), retaining the largest connected component, with the lattice
side chosen to reach $N\approx 140$, $160$, $180$, and $200$ (ten instances
per bin), solved and timed with the same code.

Figure~\ref{fig:sm_classical}(a) plots the measured $\STS(r{=}1)$
directly against the exact-solve time, one point per size group. The shot
count outgrows the solver time: the former is exponential in~$N$ while the
latter is subexponential [Fig.~\ref{fig:sm_classical}(c)], so the relation
curves upward on the log--log axes. The per-instance shell model (open
symbols) reproduces the growth but underestimates the shot count by up to a
factor of $\approx 3$; as discussed in Sec.~\ref{subsec:validation}, this deviation is
confined to the exact target.

This gap does not signal a failure of the shell model. The fitted
distribution reproduces the measured one closely: across the $96$ annealing
instances of bins~1--4 the median $D_{\mathrm{KL}}$ is $0.05$, with $78\%$
below the $0.1$ threshold of Sec.~\ref{subsec:validation} (for a representative
xlarge instance the fit matches every shell $j\ge 1$ to within a few counts
out of $500$, while $j=0$ is missed by $\sim 30\%$). The maximum-likelihood
fit is dominated by the bulk shells, and no single value of~$\beta$ in
Eq.~\eqref{eq:boltz_shell} can match both the bulk and the $j=0$ population:
refitting~$\beta$ to reproduce the measured $\pi_0$ shifts it by only
$\approx 0.14$ but then misfits the first shell by $\sim 30\%$. The residual
is concentrated at $j=0$ because $d_{\alpha}$ counts all maximum independent
sets whereas the pipeline reaches only the $1$-swap-stable ones
(Sec.~\ref{sec:discussion}). The offset is systematic rather than
statistical---the per-instance direct-to-model ratio exceeds one for
$87$--$100\%$ of instances---which is why it is not covered by the
instance-to-instance error bars. Equation~\eqref{eq:boltz_shell} therefore
remains accurate everywhere except at the exact-hit probability, where
shell-model $\STS(r{=}1)$ is a lower bound (Sec.~\ref{subsec:validation}).

Figure~\ref{fig:sm_classical}(b) applies the
same comparison at the relaxed cutoff $J(0.9)$: both estimates are
$\mathcal{O}(1)$ and agree to within about $25\%$ (direct-count medians
$1.1$--$1.7$ shots, shell-model medians $1.6$--$2.5$), while the classical
reference---one pass of the greedy pipeline---costs only
$0.05$--$1.6~\mathrm{ms}$ per instance.

\clearpage


\end{document}